\documentclass[a4paper]{jpconf}
\usepackage{iopams}
\usepackage{graphicx}
\usepackage{sidecap}
\usepackage{epsfig}
\usepackage{epsf}
\usepackage{graphicx,amsmath}
\def\beq{\begin{equation}}
\def\eeq{\end{equation}}
\def\bea{\begin{eqnarray}}
\def\eea{\end{eqnarray}}
\def\Tobs{T_{\textrm{\mbox{\tiny{obs}}}}}
\def\Tcoh{T_{\textrm{\mbox{\tiny{coh}}}}}

\def\sci#1#2{#1\times10^{#2}}

\begin{document}
\title{Hierarchical Hough all-sky search for periodic gravitational waves in LIGO S5 data}
\author{Llucia Sancho de la Jordana for the LIGO Scientific Collaboration  and the Virgo Collaboration}
\address{Departament de F\'{i}sica, Universitat de les Illes Balears, Cra. Valldemossa Km. 7.5, E-07122 Palma de Mallorca, Spain}
\ead{llucia.sancho@uib.es}

\begin{abstract}
We describe a new pipeline used to analyze the data from the fifth science run (S5) of the LIGO detectors 
to search for continuous gravitational waves  from isolated 
spinning neutron stars. The method employed is based on  the Hough transform, which is a 
semi-coherent, computationally efficient, 
and robust pattern recognition technique. The Hough transform  is used to find signals in the 
time-frequency plane of the data whose frequency evolution fits the pattern produced by 
the Doppler shift imposed on the signal by the Earth's motion and the pulsar's spin-down 
during the observation period. The main differences with respect to previous Hough 
all-sky searches are described. These differences include
the use of a two-step hierarchical Hough search, analysis of coincidences  
among the candidates produced in the first
and second year of S5, and veto strategies based on a $\chi^2$ test.

\end{abstract}

\section{Introduction}
\label{sec:intro}

Spinning neutron stars are the most promising sources of continuous gravitational wave signals
for ground-based interferometers such as GEO600~\cite{Luck:2006ug}, LIGO~\cite{Abbott:2007kv}
 and VIRGO~\cite{Acernese:2008zza}.
Using data from the different science runs, there have been two kinds of 
searches for gravitational waves from pulsars:
 i) targeted searches~\cite{Abbott:2003yq,Abbott:2004ig,Abbott:2007ce,Abbott:2008fx,Collaboration:2009rf} 
 for periodic gravitational  radiation from pulsars whose parameters 
 (sky-position and frequency evolution) are known through radio observations,
 and ii) searches for pulsars yet unobserved by radio 
 telescopes~\cite{Abbott:2006vg,Abbott:2005pu,:2007tda, Abbott:2008uq, Collaboration:2009nc, :2008rg}. 
 For this second kind of search the optimal method based on a coherent integration over the 
 full observation time is computationally prohibitive for a wide-parameter search. Therefore  all-sky
 searches require the use of semi-coherent 
 techniques~\cite{Krishnan:2004sv,Palomba:2005fp,Sintes:2006uc,
 Antonucci:2008jp,Brady:1998nj,:2007tda},
 that are less sensitive for the same observation time but are computationally inexpensive, 
 or hierarchical approaches that combine both methods, optimizing the sensitivity of a search for a given
 computational power~\cite{Papa:2000wg, Brady:1998nj,Frasca:2005ey, Cutler:2005pn,Collaboration:2009nc}. 
 Some of these semi-coherent methods operate  on successive Short Fourier Transforms (SFTs) of the measured 
 strain~\cite{Astone:2005fj} searching for cumulative excess power from hypothetical 
 periodic gravitational wave signal taking into account the Doppler modulation of the 
 detected frequency due to the Earth's rotational and orbital motion with respect 
 to the Solar System Barycenter, 
 and the time derivative of the frequency intrinsic to the source. 
 The Hough transform~\cite{Abbott:2005pu,:2007tda, Krishnan:2004sv, Palomba:2005fp} is an example of 
 such a method. 

Two flavors of the Hough transform have been developed and employed for different searches. 
The `standard Hough'~\cite{Abbott:2005pu,Krishnan:2004sv}, computes the cumulative 
excess power as a sum of binary zeroes and ones, where a SFT contributes a one if and only 
if the power exceeds a normalized power threshold, and the
 `weighted Hough'~\cite{:2007tda, Sintes:2006uc} in which the contribution of each SFT is 
 weighted according to the noise and detector antenna pattern to maximize the signal-to-noise ratio. 
 Using the `weighted Hough' has two notable benefits: i) 
  the gain in sensitivity, and ii)
 that this method can be used to analyze data from multiple detectors, since
 the different weights automatically take into account the different sensitivities.

 This paper describes the pipeline employed to analyze the data from the fifth science run of the LIGO
 detectors to search for periodic gravitational waves using a hierarchical Hough transform, emphasizing 
 the changes with respect to previous Hough searches~\cite{Abbott:2005pu, :2007tda}. 
 The paper is organized as follows:
 In section~\ref{sec:Hough} we briefly summarize the basic principles of the Hough transform and its statistics. 
 In section~\ref{sec:pipeline}  we give an overview of the search pipeline and comment on the data employed 
 and the parameter space of the search. In section~\ref{sec:improvements} we describe two main 
 features of our two-step hierarchical search, and we conclude in section~\ref{sec:conclusions}.

\section{The Hough transform}
\label{sec:Hough}

The starting point for the Hough transform is a set of $N$ SFTs, short stretches of Fourier transformed data
which are digitized by setting a threshold $\rho_{th}$ on the normalized power 
$\rho_k$ in the different frequency bins: 
\begin{equation} 
\label{eq:1}
  n_k^{(i)} = \left\{ 
  \begin{array}{ccc} 
    1  & \textrm{if} & \rho_k^{(i)} \geq \rho_{th}  \\
    0  & \textrm{if} & \rho_k^{(i)} < \rho_{th}  
  \end{array}  \right.\, ,
\end{equation}
where
\beq
\rho_k = \frac{2 \vert \tilde{x_k} \vert^2}{\Tcoh S_n(f_k)} \, .
\eeq
Here $\tilde{x_k}$ is the value of the Fourier transform in the $k^{th}$ frequency bin corresponding 
to a frequency $f_k$, $\Tcoh$ is the time baseline of the SFT (in our case $\Tcoh=1800~\mathrm{s}$), 
and $S_n(f_k)$ is the single sided power spectral density of the detector noise at the frequency $f_k$. 

For a given template, the Hough number count is a weighted sum of the binary zeroes and ones, along the corresponding
 time-frequency pattern
\begin{equation}
\label{eq:2}
  n = \sum_{i=0}^{N-1} w^{(i)}_{k} n^{(i)}_{k}\, ,
\end{equation}
where the weights are defined as
\begin{equation} 
\label{eq:3}
  w^{(i)}_{k} \propto  \frac{1}{S^{(i)}_{k}}\left\{
    \left(F_{+1/2}^{(i)}\right)^2 +
    \left(F_{\times 1/2}^{(i)}\right)^2\right\} ,
\end{equation}
being $F_{+1/2}^{(i)}$ and $F_{\times 1/2}^{(i)}$ the values of the beam pattern functions at the mid point of the $i^{th}$ SFT. 
Thus, we take a binary count $n^{(i)}_{k}$ to have greater weight if SFT $i$ has a lower noise floor or if, 
in the time-interval corresponding to this SFT, the beam pattern functions are larger for a particular point in the sky.  

The weight normalization is chosen according to
\begin{equation}
 \label{eq:4}
  \sum_{i=0}^{N-1} w^{(i)}_{k} = N\, .
\end{equation}
With this normalization the Hough number count $n$ lies within the range $[0,N]$. 
Note that the sensitivity of the search is governed by the ratios of the different weights, not by  the choice of overall scale.
The robustness of the Hough transform method in the presence of strong transient disturbances is not compromised by using weights 
because each SFT contributes at most $w_i$ (which is of order unity) to the final number count. 

The natural detection statistic is not the Hough number count $n$, but the \textit{significance} of a number 
count, defined by:
\beq
s=\frac{n-\bar{n}}{\sigma} \, ,
\eeq
where $\bar{n}$ and $\sigma$ are the expected mean and standard deviation for pure noise. 
Values of $s$ can be compared directly across different templates characterized by different weight distributions and $\sigma$  values.
Furthermore, in the case of Gaussian noise, the relation between the significance and the false alarm probability $\alpha$ is given by:
\beq
\alpha = 0.5 \,\textrm{erfc} (s/\sqrt{2}) \, .
\eeq
Setting a threshold on the significance would then identify interesting candidates. 
We refer the reader to~\cite{:2007tda} for further details.


\section{Description of the pipeline}
\label{sec:pipeline}

This paper presents a new pipeline used to analyze the data from 
the LIGO detectors to search for continuous gravitational waves from 
isolated spinning neutron stars.
The LIGO detector network consists of two interferometers in 
Hanford Washington, one 4-km and another 2-km (H1 and H2) and 
a 4-km interferometer in Livingston Louisiana (L1).
The search described here is currently carried out over the entire sky using 
 the data produced during LIGO's fifth science run (S5) that started on November 4, 2005 
and ended on October 1, 2007, at initial LIGO's design sensitivity.
Data from each of the three LIGO interferometers is used to perform the
 search.
 
 The starting point is a collection of SFTs generated  directly from the calibrated data stream, 
 using 30-minute intervals of data for which the interferometer is operating in what is
 known as science mode. With this requirement, we search 32295 SFTs from the first year of S5
 (11402 from H1, 12195 from H2 and 8698 from L1) and 35401 SFTs from the second year 
 (12590 from H1, 12178 from H2 and 10633 from L1). 
 
 The search is performed in the frequency range  $50\,$--$\,1000$~Hz
 and with the frequency's time derivative in the range 
$-\sci{8.9}{-10}~\mathrm{Hz}~\mathrm{s}^{-1}$ to zero, being those values limited
by the computational cost of the search. We use a uniform grid spacing equal to the
size of a SFT frequency bin, $\delta f = 1/ \Tcoh= \sci{5.556}{-4}~\mathrm{Hz}$.
The resolution $\delta \dot{f}$  is given by the smallest value of $\dot{f}$ for which 
the intrinsic signal frequency does not drift by more than a  frequency bin during 
the total observation time $\Tobs$: 
$\delta \dot{f} = \delta f / \Tobs \sim \sci{1.8}{-11}~\mathrm{Hz}~\mathrm{s}^{-1}$.
This yields 51 spin-down values for each frequency. 
$\delta \dot{f}$  is fixed to the same value for the search on the first and the second year
 of S5 data, being $\Tobs$ the value for the first year.
The sky resolution, $\delta \theta$, is frequency dependent, 
with the number of templates increasing with frequency, as given by 
Eq.(4.14) of Ref.~\cite{Krishnan:2004sv}. This yields a resolution of about 
$9.3 \times 10^{-3}$~rad at $300$~Hz, which corresponds to 
$\sim 1.5 \times 10^5$ sky locations for the whole sky at that frequency.

The key difference from previous searches is that, starting from 
$30~\mathrm{min}$ SFTs, 
we perform a multi-interferometer search analyzing separately the 
two years of the S5 run, and we study coincidences among the source candidates 
produced by the first and second years of data. This 
is inspired by a similar coincidence search using VIRGO data~\cite{Acernese:2007zzb}.
Furthermore, we use a $\chi^2$ test adapted to the Hough transform searches,
as described in~\cite{SanchodelaJordana:2008dc}, to veto potential candidates.

The approach used to analyze each year of data is based on a two-step hierarchical search
for continuous signals from isolated neutron stars as described below.
In both steps, the weighted Hough transform is used to find signals whose frequency 
evolution fits the pattern produced by the Doppler shift and the spin-down in the 
time-frequency plane of the data. The search is done by splitting the 
frequency range in $0.25$~Hz bands and using the SFTs from multiple interferometers. 

In the first stage, 
we break up the sky into smaller patches with frequency dependent size.
The size of the sky-patches ranges from $\sim 0.4~\mathrm{rad} \times 0.4~\mathrm{rad}$
at $50$~Hz to $\sim 0.07 ~\mathrm{rad} \times 0.07~\mathrm{rad}$ at $1$~kHz.
Ideally, to obtain the maximum increase in sensitivity, we should calculate the 
weights, based both on the noise and the antenna pattern, for each sky-location. 
In practice,  we calculate the weights just once for the center of each sky-patch. 
In this first stage, we perform the Hough multi-interferometer search using the traditional 
\textit{look up table} approach, that enormously reduces the computational cost 
(see~\cite{Krishnan:2004sv} for a detailed description). 
But limitations on the memory of the machines
constrain the volume of data (i.e., the number of SFTs) that can be analyzed at once 
and the parameter space (e.g., size and resolution of the sky-patches and number 
of spin-down values) we can search over.
For this reason, in this first stage, the Hough transform is not applied  
using all the available SFTs, but selecting the best 15000 SFTs for each 
frequency band and sky-patch.
 A top-list keeping the best $1000$ candidates is produced for each $0.25$~Hz band. 

In a second stage, we recompute 
the Hough significance of each candidate in the top-list using 
the complete set of available 
SFTs from all the interferometers, and at the same time we reduce
 the mismatch of the 
template, i.e., we calculate the number count and the corresponding significance
without the roundings introduced 
by the \textit{look up table} approach and by recomputing the  
weights for the precise sky location and not for the center of the corresponding patch, as
previously done. 
For each candidate, we also compute the $\chi^2$ value that will help in vetoing 
the resulting candidates.

\section{A two-step hierarchical Hough search}
\label{sec:improvements}

In this section we describe in more detail the two main features of the
 two-step hierarchical Hough search, i.e., the selection of the best SFT data
 and the comparison of the significance values produced in both steps.
 
\subsection{Selection of the best SFT data}
\label{subsub:best}

The first of the two main changes to the Hough code consists 
in selecting the best SFTs based on the weights given by Eq.~(\ref{eq:3}). 
In this way, when doing the all-sky search for each 0.25~Hz band, 
for each sky-patch, we will keep the 15000 SFTs
 that have lower noise and that are more sensitive at that sky location.
 
\begin{figure}[h]
\begin{minipage}{14pc}
\includegraphics[width=20pc]{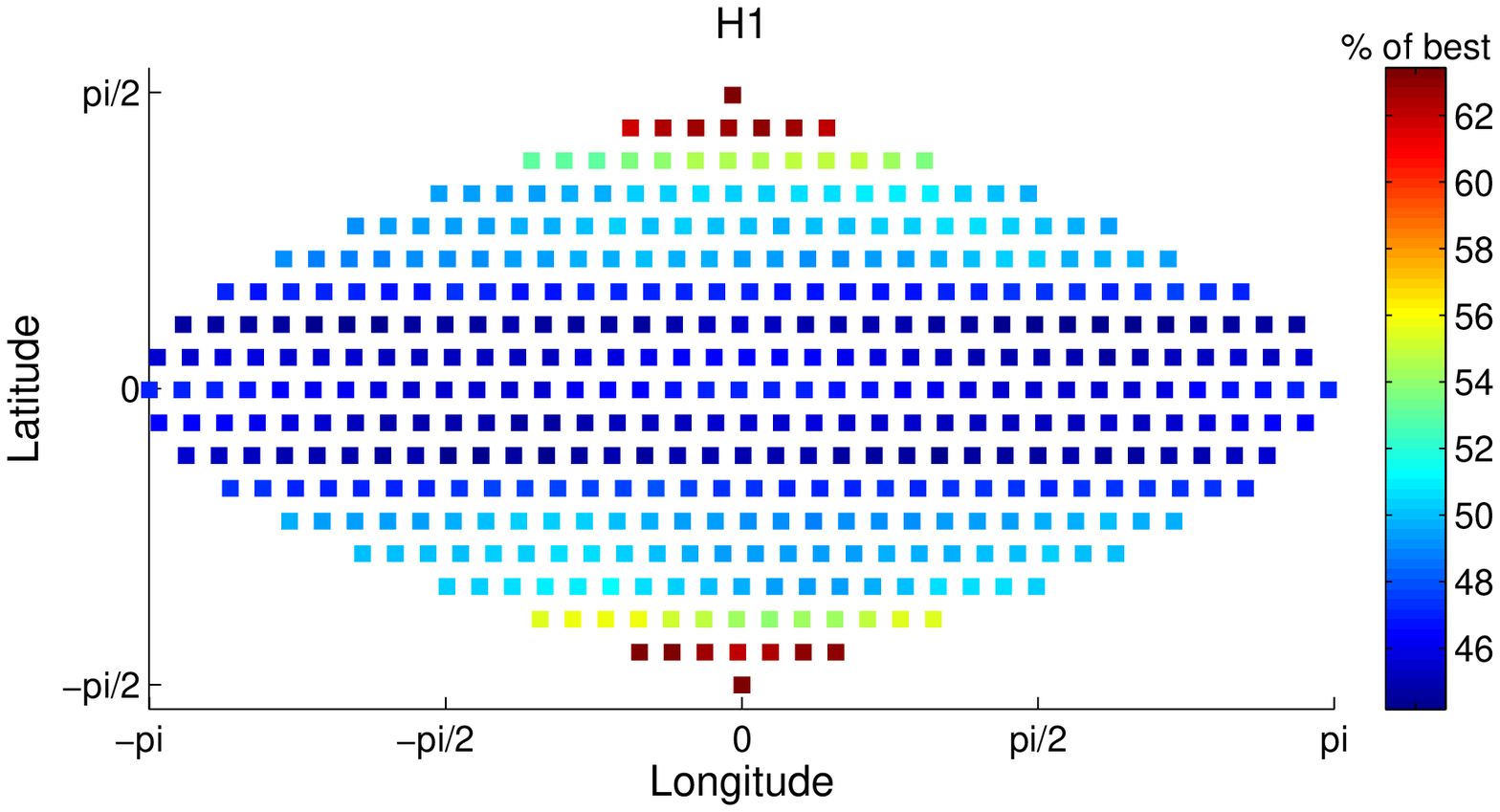}
\end{minipage}\hspace{6.5pc}
\begin{minipage}{14pc}
\includegraphics[width=20pc]{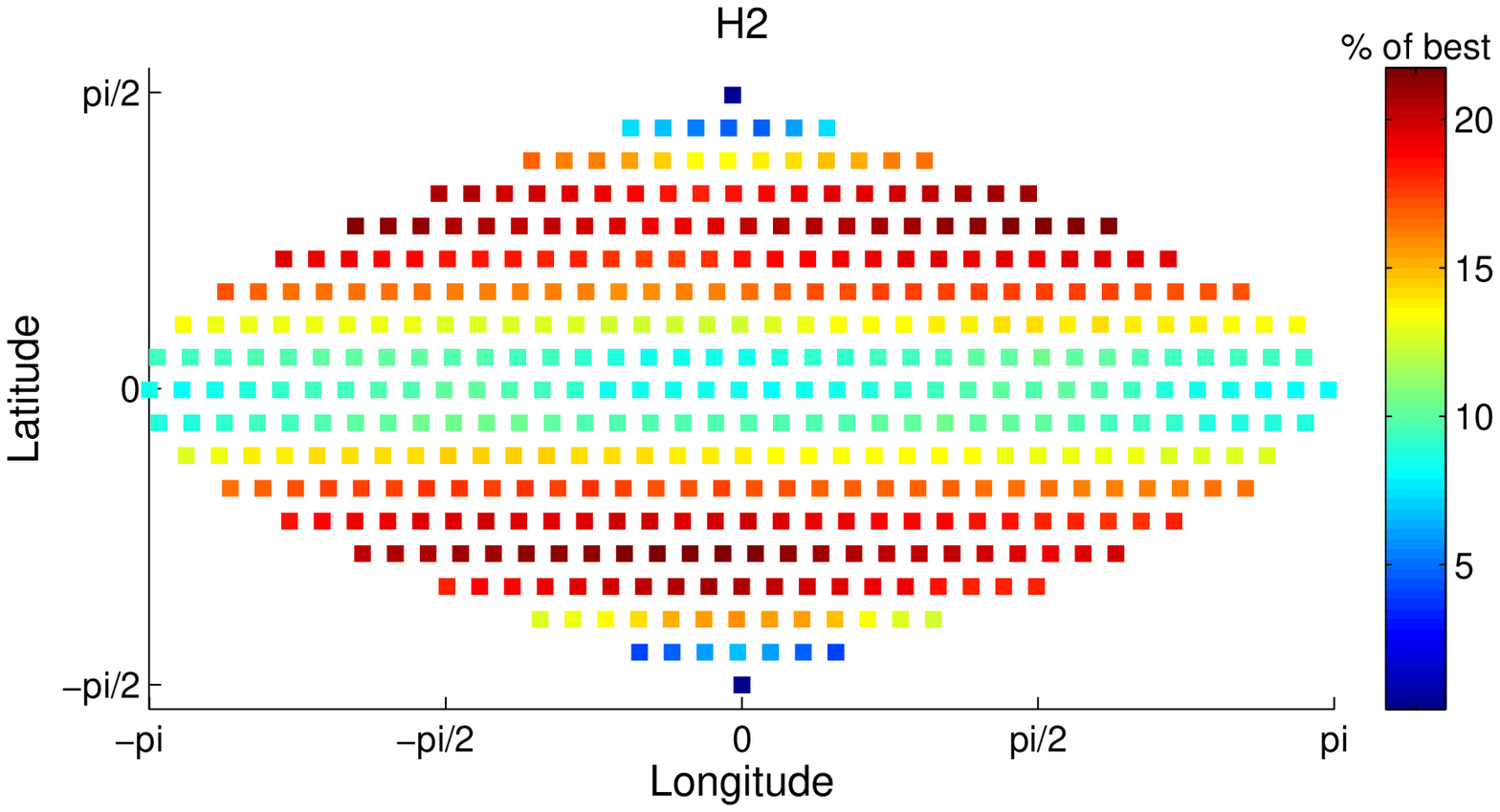}
\end{minipage}

\begin{minipage}{14pc}
\vspace{-1pc}
\includegraphics[width=20pc]{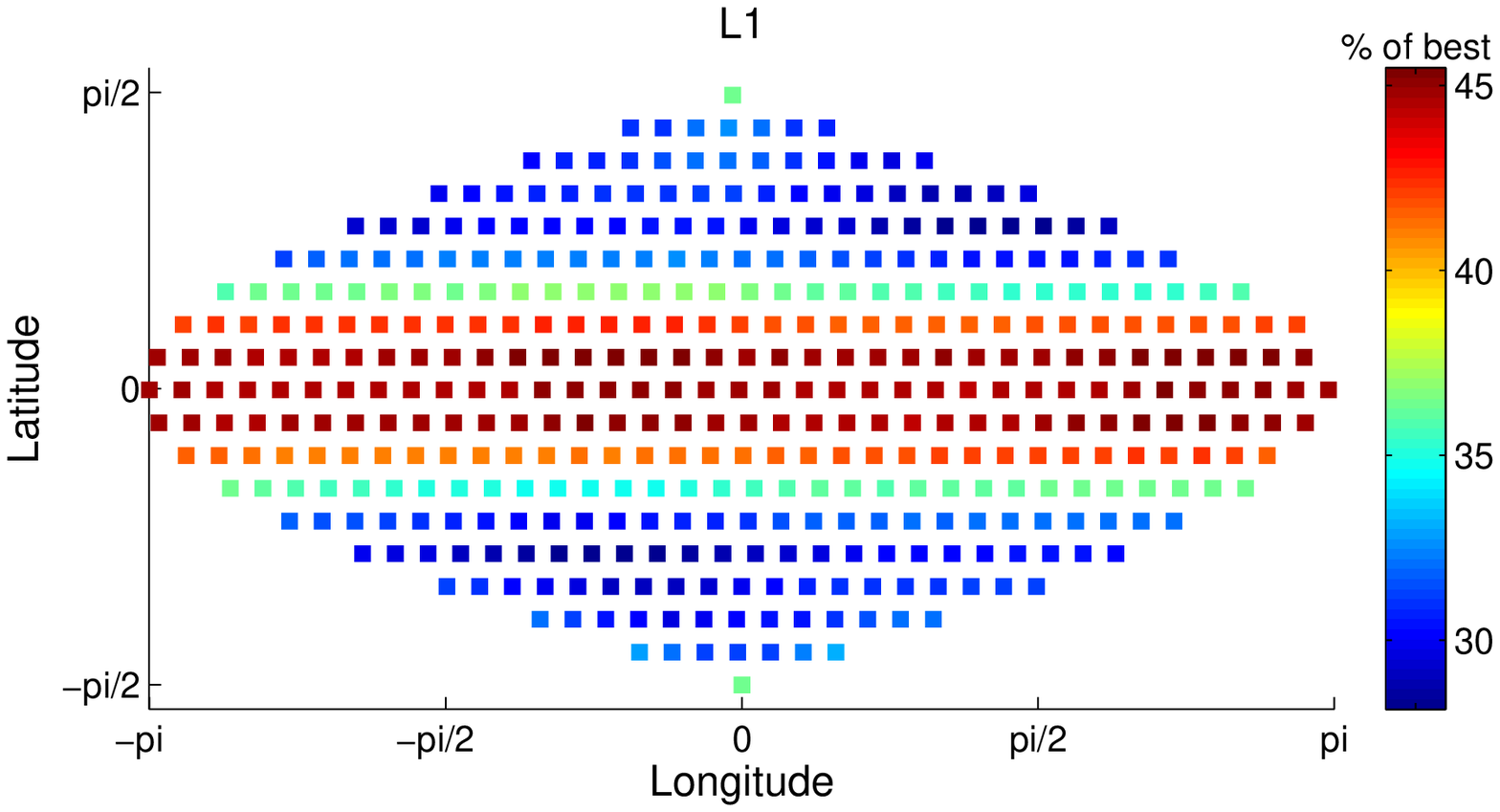}\hspace{2pc}%
\end{minipage}\hspace{7pc}
\begin{minipage}[b]{19pc}
\caption{\label{label} Percentage of SFTs 
that each detector has contributed at each particular sky-patch when selecting the best 15000 SFTs
 based on the Hough weights. 
 These figures correspond to an all-sky search performed on a $0.25$~Hz
  band at $420$~Hz for the S5 first year of data. }
\label{fig:1}
\end{minipage}
\end{figure}

Figure \ref{fig:1} shows the percentage of SFTs from each of the 
three LIGO detectors (H1, H2 and L1) that have been selected when doing 
the Hough search on a $0.25$~Hz band at a frequency of 420~Hz for the 
S5 first year of data.  At this frequency, the number of sky-patches  is
 $426$ with a size of $\sim 0.17 ~\mathrm{rad} \times 0.17 ~\mathrm{rad}$. 
 The figure shows the percentage of SFTs from each detector on each sky-patch. 
 In this figure we can see that, for this particular frequency, the detector that contributes 
 the most at almost all the sky locations is H1. The maximum contribution of H1, about $64\%$, 
 is at the poles, L1 has it maximum contribution, about $46\%$, around the equator 
 and H2 contributes at most $23\%$ of the SFTs in the region in between.

The selection of SFTs has been done based on the total weights,
that depend both on noise floor and antenna pattern. If this selection had been based
only upon the weights due to the noise floor, 
the H2 detector would not have contributed at all in this first stage. 
By having the weights antenna pattern dependent, H2 has a certain contribution 
in some 
sky locations where it is more sensitive.

\subsection{Comparison of the significance values}
\label{subsub:followup}

In the first stage of the hierarchical search, the Hough 
significance is computed using the \textit{look up table} approach 
to enormously reduce the computational cost.
The code uses the selected $15000$ best SFTs to compute the Hough significance 
and make a list of interesting candidates.

In the second stage of the search, for each candidate stored in the top-list 
produced for each $0.25$~Hz band, the code recomputes the Hough significance 
without any rounding, and using all the SFTs from the three LIGO detectors. 
In this stage, the significance is computed directly from the number count that is 
obtained from the digitized time-frequency plane by summing weighted binary 
zeros or ones along the expected path of the frequency of a hypothetical 
periodic gravitational wave signal, taking into account the Doppler shift and 
the spin-down.

\begin{figure}[h]
\begin{minipage}{14pc}
\includegraphics[width=20pc]{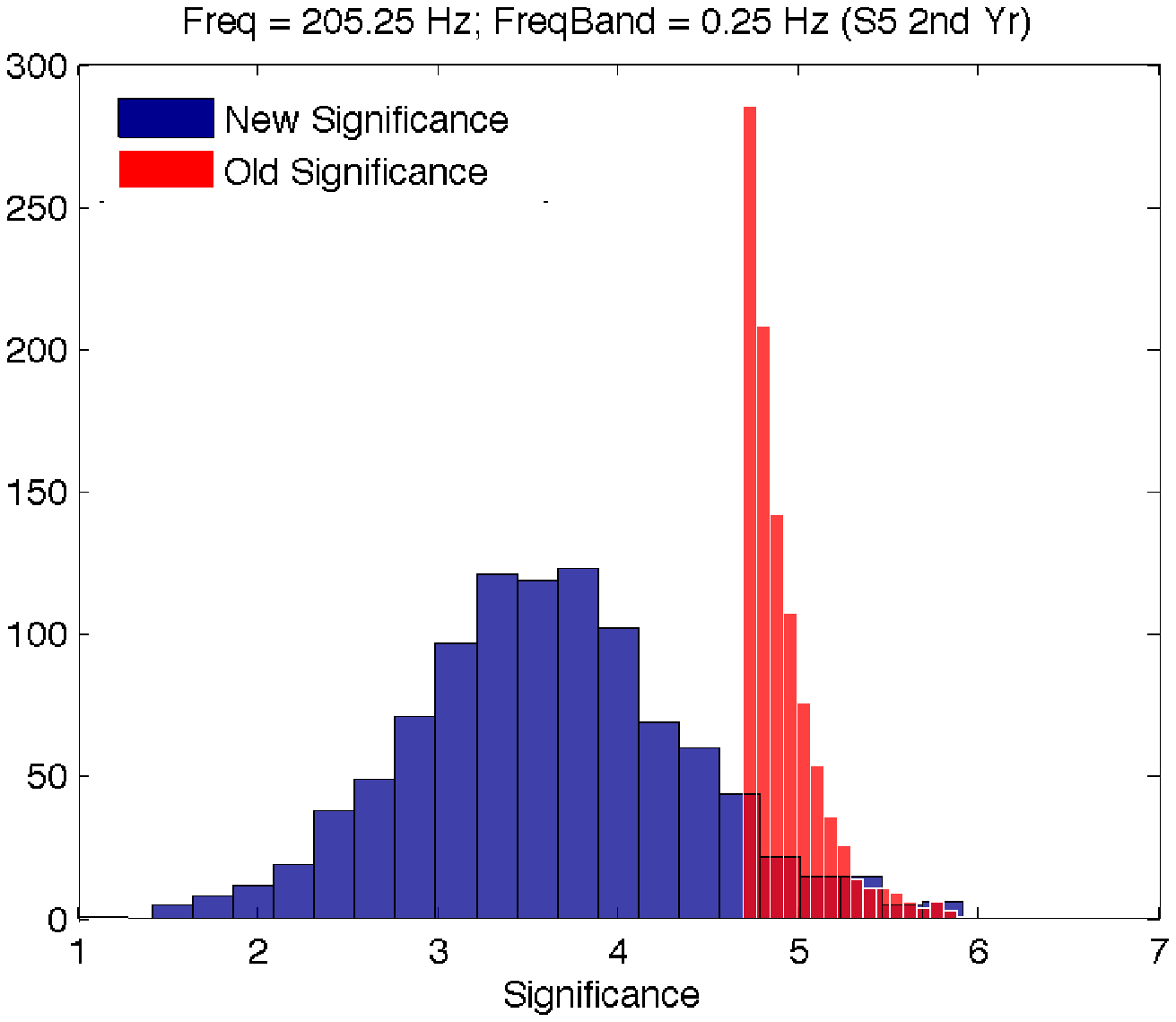}
\end{minipage}\hspace{5pc}
\begin{minipage}{14pc}
\includegraphics[width=20pc]{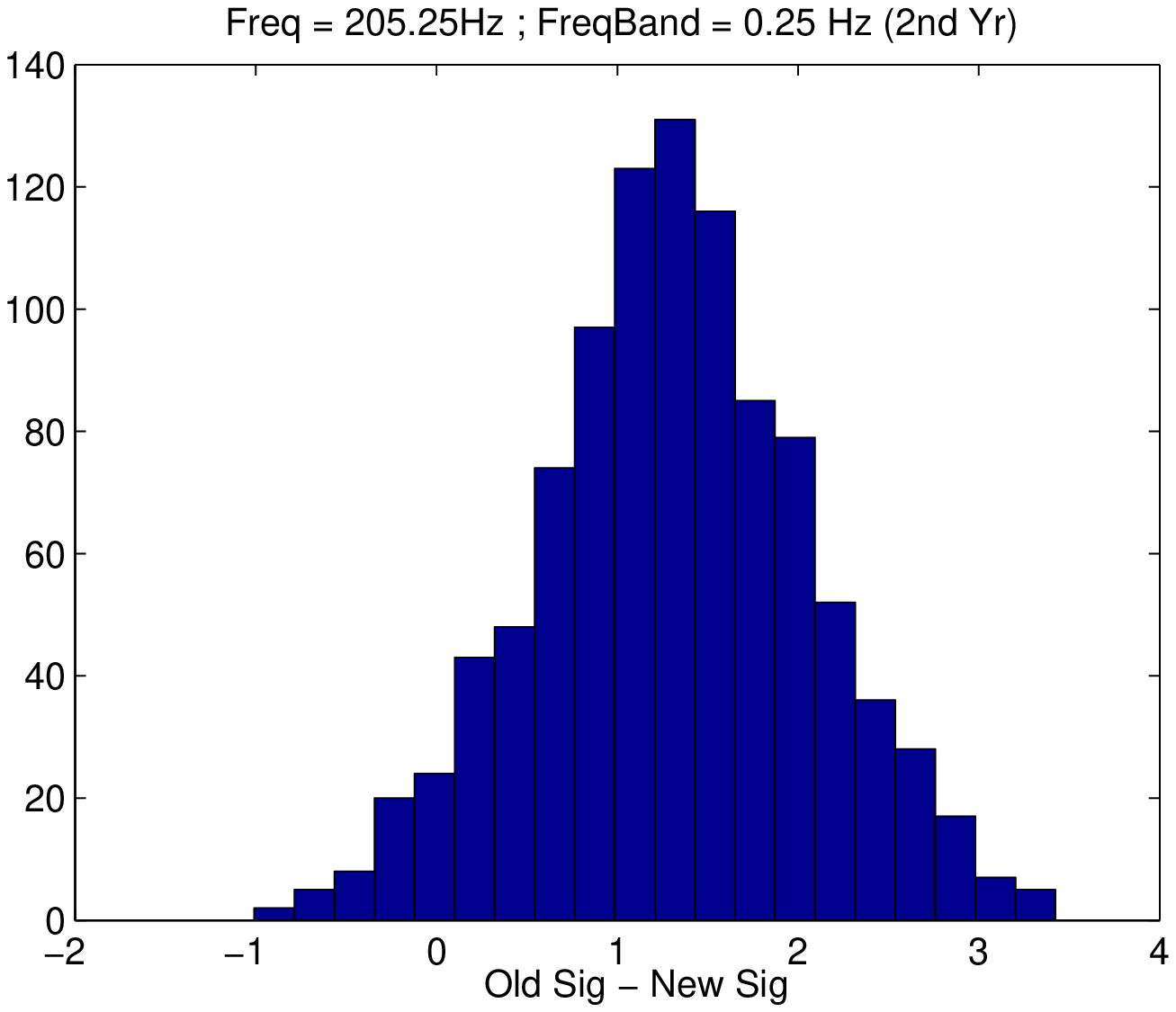}
\end{minipage}

\begin{minipage}{14pc}
\includegraphics[width=20pc]{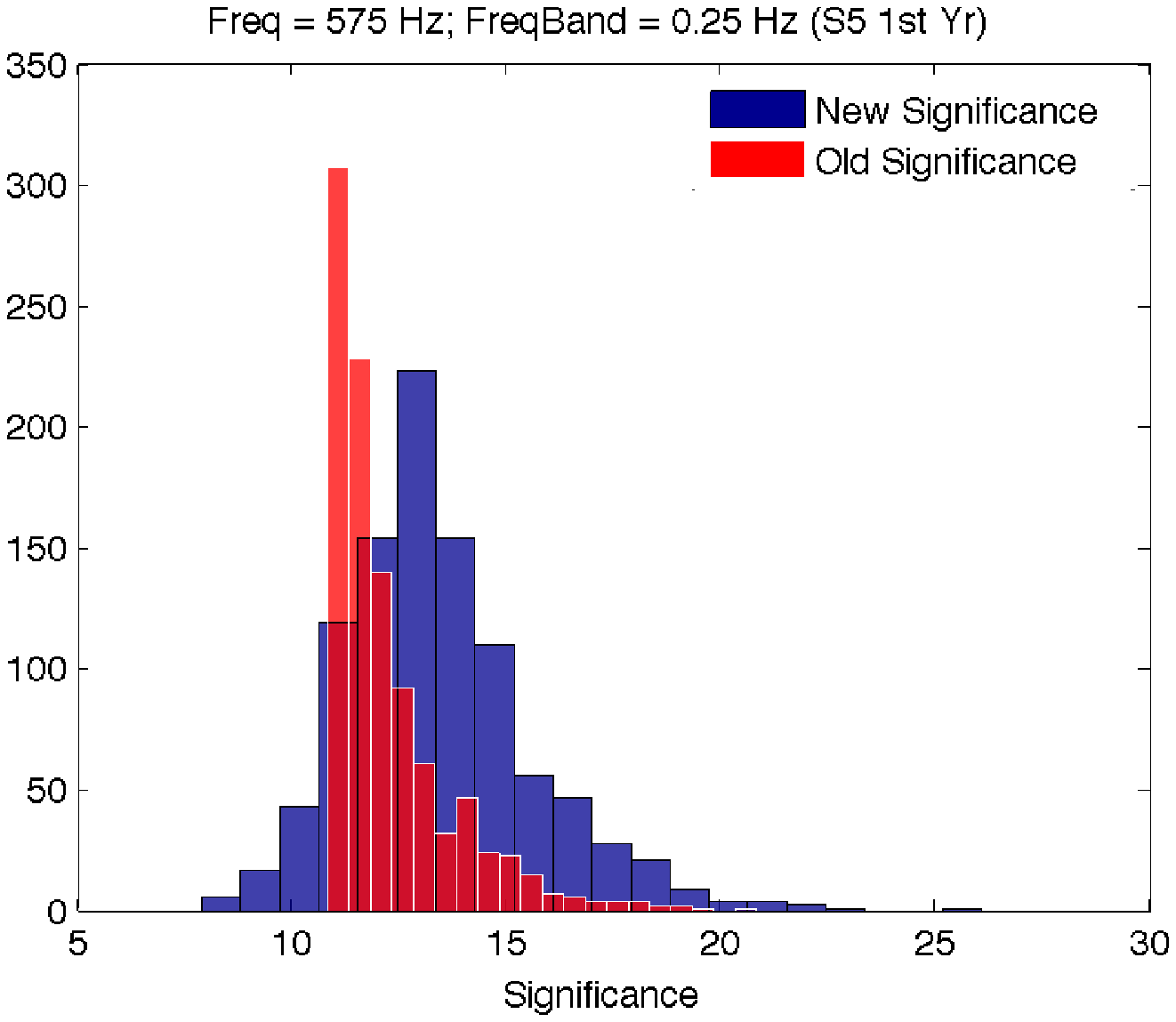}\hspace{2pc}%
\end{minipage}\hspace{5pc}
\begin{minipage}{14pc}
\includegraphics[width=20pc]{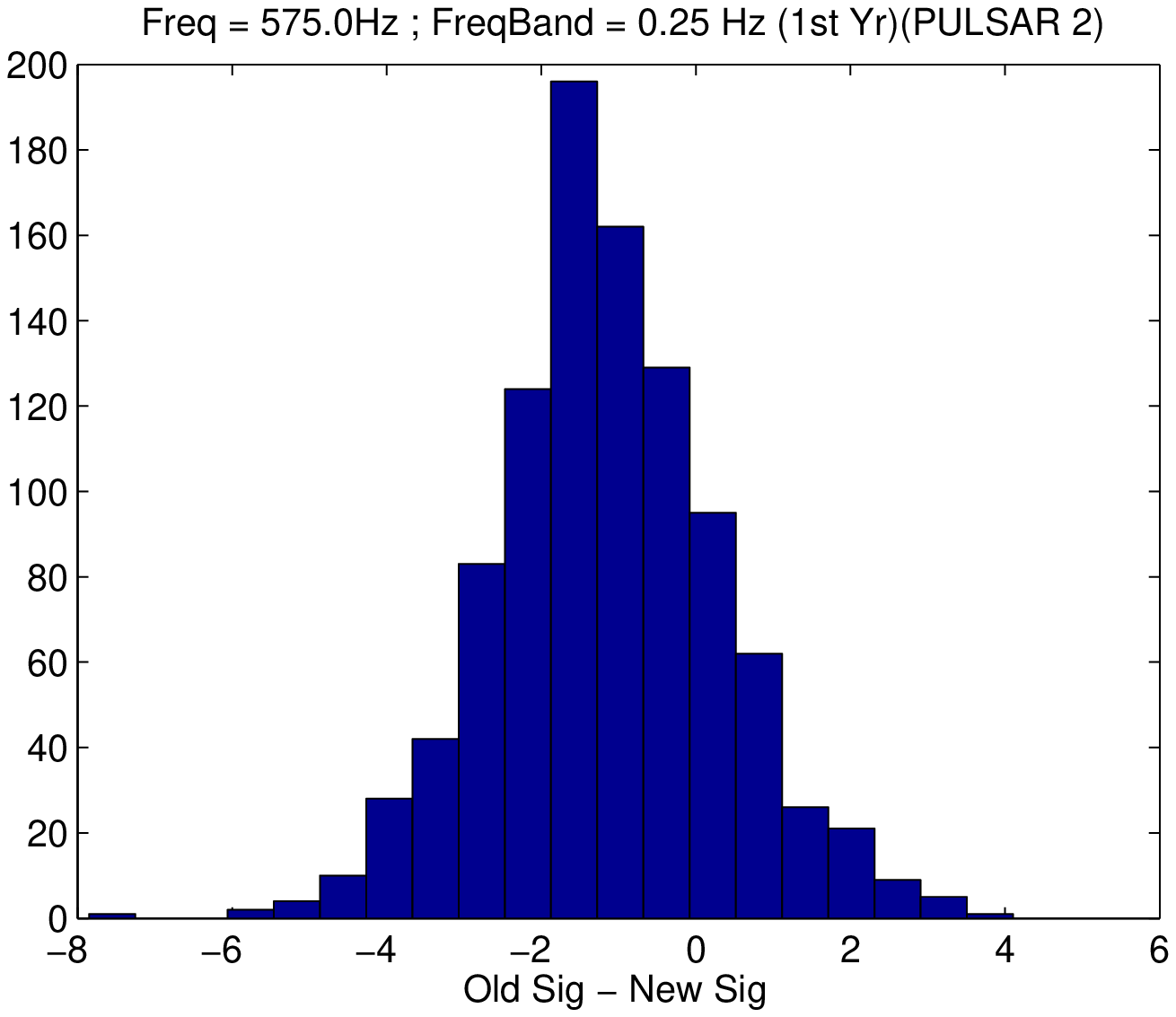}\hspace{2pc}%
\end{minipage}

\begin{minipage}{38pc}
\caption{\label{label} Comparison of the distribution of the significance 
values when calculated in the first stage using the \textit{look up table} 
approach (`old significance') and after recomputing its value using all the 
SFTs and without any rounding (`new significance'). 
These results correspond to an all-sky search  on a $0.25$~Hz
band at a frequency of $205.25$~Hz for the S5 2nd year of data (top panels), where no 
relevant spectral disturbances were present, 
and  at $575$~Hz for the 1st year of data (bottom panels), containing 
a hardware injected pulsar. 
 }
\label{fig:2}
\end{minipage}
\end{figure}

Figure \ref{fig:2} shows how this recomputation affects the previous 
value of the significance in the case where an injected pulsar was present and 
in the case of a frequency band free of spectral disturbances.
To illustrate the effect, we have used a `quiet band' 
at $205.25$~Hz and a band with an injected pulsar at $575$~Hz. In both bands we
 perform an all-sky 
search and compare the significances computed on the first 
and second stages. The `old significance' corresponds to 
the one computed in the first stage, using the \textit{look up table} 
approach and using only the best 15000 SFTs. 
The `new significance' corresponds to the 
one computed in the second stage using all the 
SFTs and without any rounding. In the same figure we also provide
 the histograms 
of the differences between the `old' and `new' significances, showing that in the 
case where we have the hardware injected signal, the values of the `new' 
significance are, in most cases, higher than the `old' values. While in 
the case of the `quiet' band, the `new significance' becomes smaller than 
the `old' one in the majority of the cases.

\section{Conclusions and future work}
\label{sec:conclusions}

In this paper, we have presented the improved Hough search pipeline, which is being used to analyze 
data from the fifth science run of the LIGO interferometers, describing in detail the two 
 main  features of the two-step hierarchical search, i.e., the selection of SFT data and 
 recomputation of the significance values.
We have shown how these improvements perform on the data, showing examples at some particular frequencies,
either with interesting artifacts, 
such as hardware injected pulsars, or others where we expect approximately Gaussian noise.
Work is in progress  to analyze the coincidences among the candidates produced in the first 
and second year of S5 for the full parameter space,
and to compute astrophysical upper limits using the search pipeline 
presented in this paper. 

\ack
The authors gratefully acknowledge the support of the United States National Science Foundation for the construction and operation of the LIGO Laboratory, the Science and Technology Facilities Council of the United Kingdom, the Max-Planck-Society, and the State of Niedersachsen/Germany for support of the construction and operation of the GEO600 detector, and the Italian Istituto Nazionale di Fisica Nucleare and the French Centre National de la Recherche Scientiﬁque for the construction and operation of the Virgo detector. The authors also gratefully acknowledge the support of the research by these agencies and by the Australian Research Council, the Council of Scientiﬁc and Industrial Research of India, the Istituto Nazionale di Fisica Nucleare of Italy, the Spanish Ministerio de Educaci´on y Ciencia, the Conselleria d’Economia Hisenda i Innovaci´o of the Govern de les Illes Balears, the Foundation for Fundamental Research on Matter supported by the Netherlands Organisation for Scientiﬁc Research, the Royal Society, the Scottish Funding Council, the Polish Ministry of Science and Higher Education, the FOCUS Programme of Foundation for Polish Science, the Scottish Universities Physics Alliance, The National Aeronautics and Space Administration, the Carnegie Trust, the Leverhulme Trust, the David and Lucile Packard Foundation, the Research Corporation, and the Alfred P. Sloan Foundation. LIGO Document No. LIGO-P0900212

\section{References}
\bibliography{Sancho_Amaldi_proceedings}

\end{document}